\documentstyle[twocolumn,prl,aps]{revtex}   

\begin{document}
\newcommand {\CC}{{\cal C}}
\newcommand {\CCD}{{{\cal C}^\dagger}}
\newcommand {\HH}{{\cal H}}

\title
{RELAXATION TIME FOR A DIMER COVERING \\
WITH HEIGHT REPRESENTATION}

\author{Christopher L. Henley}

\address{Dept. of Physics, Cornell University,
Ithaca NY 14853-2501}
\maketitle

\begin{abstract}

This paper considers the Monte Carlo dynamics of random dimer coverings
of the square lattice, which can be mapped to a rough interface model.
Two kinds of slow modes 
are identified, associated respectively 
with long-wavelength fluctuations of the interface height, 
and with slow drift (in time) of the system-wide mean height.
Within a continuum theory, the longest relaxation time 
for either kind of mode scales as the system size $N$.
For the real, discrete model, an exact {\it lower} bound of $O(N)$ is placed 
on the relaxation time, 
using variational eigenfunctions corresponding to the
two kinds of continuum modes.

\end{abstract}
%
%
%
%
\narrowtext  
%
%

\def \half {{\scriptstyle{1\over 2}}}
\def \quart{{\scriptstyle{1\over 4}}}
\def \hbar {{\overline h}}
\def \zbar {{\overline z}}
\def \W {{\cal W}}
\def \P {{\cal T}}
\def \T {{\cal T}}
\def \tz {{\tilde z}}
\def \th {{\tilde h}}
\def \x {{\bf x}}
\def \r {{\bf r}}
\def \R {{\bf R}}
\def \q {{\bf q}}
\def \lambdahbar {{\lambda_{\hbar}}}

\section{INTRODUCTION}
\label{intro}

Statistical models in which each microstate maps to an interface --
which I will call ``height models'' -- tend to be interesting, for
the interface is often rough, and it turns out the model has
critical correlations.~\cite{blot,zheng,lev,thurst,kond,rag,zeng,burt}.
This paper considers the corresponding {\it dynamics}, 
using one of the simplest lattice statistical models: the dimer covering
of the square lattice\cite{fish,kast}.
The dimers are placed on the bonds and
every site must be touched by exactly one dimer.
(This is equivalent, of course, to the packings of ``dominoes''
of size $2 \times 1$ \cite{xpropp}; furthermore the ground
states of the
fully frustrated Ising model on the square lattice~\cite{vill}
map 2-to-1 to the dimer coverings. 
In this paper, I only consider the case where each 
packing has equal weight.
The system is taken to be a square of $L\times L$ sites with 
periodic boundary conditions for the dimers. 

The statics can be described by mappings to free fermions,
and exact solutions are possoble using Pfaffians
\cite{fish,kast} or transfer matrices \cite{lieb,suth}.
There is a nonzero ground state entropy of 0.2916 per site
\cite{fish}.
The correlation functions are critical~\cite{fishs} (power-law decaying).
This is easiest understood after mapping the dimer 
packings (configuration-by-configuration) to configurations 
of ``heights'' $z(\r)$ living on dual lattice sites, 
$\r =(x,y)$, 
representing a rough interface in an 3-dimensional abstract  space
\cite{blot,zheng,lev,thurst,kond,rag,zeng,burt}.

When such a model is endowed with dynamics, a central question
is how the equilibration time $\tau(L)$ 
(to be defined shortly) scales as a function of the system diameter $L$?
In particular, what is the dynamic exponent $z$ \cite{hoh} 
in $\tau(L) \sim L^z$?
Since the static continuum model has a gradient-squared free energy, 
one would suspect that $z=2$. 
Indeed, scaling with $z=2$ has been seen numerically 
in simulations of the square lattice dimer model\cite{propp}.
(It has also been seen in antiferromagnetic Ising models\cite{zeng}
and random-tiling quasicrystal models\cite{shaw} 
with height representations.)
However, it has only been proven \cite{luby} that 
(with our definition of time scale)
$\tau(L) \leq O (N^3$), which implies $z\leq 6$.

This paper has two aims:
(i) an explicit approximate description of the slowest eigenmodes
of the time evolution, based on continuum theory -- 
this theory is applicable, with small changes, to any height model;
(ii) the outline of a proof 
that $\tau (L) \ge O(L^2)$ (and hence $z\ge 2$.)
The 

The outline of the paper is as follows.
The remainder of this section specifies the model, in particular
the height mapping and dimer-flip stochastic dynamics, and also
points out an exact correspondence of the {\it entire spectrum}
of the quantum dimer model to that of this dynamics. 
I then take up a continuum model 
in terms of the height representation
(Secs.~\ref{cont}) which is
defined by standard Langevin dynamics, verifying  that this is the
appropriate coarse-graining of the microscopics. 
This produces an approximate description of
{\it all} the slow eigenmodes (Secs.~\ref{cont} and~\ref{rwalk}). 
Intriguingly, the slowest mode in any finite-size system is {\it not} the 
longest-wavelength capillary wave, but a ``height-shift'' mode
involving diffusion 
of the system-wide average ``height'' direction (Sec.~\ref{rwalk}). 

Finally (in Sec.~\ref{micro}) 
I explain how rigorous bounds on the dynamics can be proven. 
The fundamental concepts for this proof are (i) the use of
Fourier analysis to partially diagonalize the evolution 
matrix $\W$, (ii) constructing variational ``wavefunctions''
guided by the results of the earlier sections,  and (iii)
taking advantage of the
Fourier spectrum of height fluctuations derivable from the exact 
solution of the model.
In the conclusion, I mention other spin models
to which these results should be relevant.

\subsection {Height representation}

The explicit rule for constructing a height pattern $\{ z(\r) \}$, 
given  a snapshot of the dimers, is shown in Fig. 1: 
$z(\r)-z(\r')=-3$ if there is a dimer between $\r$ and 
$\r'$, or $+1$ if there is no dimer, where the step from 
$\r$ to $\r'$ is
taken in a counterclockwise sense about the even sites.
The net height difference is zero for the path around one plaquette, 
and hence by induction for any closed path, showing the consistency
of the definition. However, it is nonunique in that 
adding the same constant to each $z(\r)$ makes an equally valid
height representation of the same dimer configuration;
I shall fix this constant after defining the dynamics (below).

It is necessary in writing the continuum model for $z(\r)$
(e.g. to define its Fourier transform) that $z(\r)$ satisfy periodic
boundary conditions; but periodic boundary conditions for dimers only imply
$z(L,y)-z(0,y)=w_x$ and $z(x,L)-z(x,0)=w_y$,
where the ``winding numbers'' $w_x$ and $w_y$  are multiples  of 4, 
so that the system has a mean tilt is $(w_x/L, w_y/L)$.
{\it Local} update rules 
(such as I am about to define) conserve the winding numbers. 
Thus one could define substracted heights 
$z'(x,y)\equiv z(x,y)-(w_x/L)x -(w_y/L)y$
which do satisfy periodic boundary conditions. 
In the limit of small tilts, 
$z'(\r)$ would obey the same continuum dynamics.\footnote{
If tilts are nonvanishing as $L\to \infty$, 
the free-energy functional (\ref{freeE})
must be generalized to have different stiffnesses for components of the
gradient parallel to and transverse to the tilt direction;
the same power laws for the dynamics would be deduced. }
This trivial generalization is not worth the added complication in
notation;  in the remaining sections I will only configurations with 
$(w_x,w_y)=(0,0)$

The ground states of the fully-frustrated
Ising model on the square lattice may be mapped (2-to-1) to 
dimer coverings of the dual square lattice. The rule is: simply draw a
dimer across every violated bond. 
Therefore that model has a height representation and
all my results apply equally to the fully-frustrated Ising model.

The height variable will play the role of the ``order parameter''
in this paper. 
The more customary order parameter, for {\it any}
of the height models, 
is a spin operator $m(\r)$
(or a dimer occupation operator, in the present case). 
However, as outlined in Appendix~\ref{spincorr}), 
such an $m(\r)$ is simply a sinusoidal function of the local height.
Since the long-distance and long-time 
height fluctuations discussed here are Gaussian, it
is possible from them to compute the correlations of $m(\r)$ 
which turn out to be algebraic with nonuniversal exponents.

\subsection {Dynamics}

It is possible to turn any dimer covering (of zero mean tilt) into any other one
by a succession of ``dimer flips''
each affecting two dimers on opposite edges of one 
plaquette
(see Fig.~1(b)).
Thus the model is endowed with a stochastic (Monte Carlo)
dynamics
in continuous time as follows:
select plaquettes at random, at a rate $N$ per unit time
where $N \equiv L^2$ is the number of sites (i.e.  on average 
visit each plaquette
once per unit time). 
Flip the plaquette if it has two dimers as in
Fig.~1(b), otherwise do nothing.

On the other hand, 
the natural zero-temperature dynamics of the fully frustrated Ising
model is to choose a spin at random and flip it if the energy change 
would be zero, otherwise do nothing. 
This induces exactly the same dynamics on the dimer configurations 
as specified in the preceding paragraph.

To eliminate the arbitrariness in defining $z(\r)$ and ensure that
the coarse-grained dynamics is continuous in time, 
we relate $z(\r,t)$ at different times,
by specifying that a dimer flip on a plaquette
changes only the $z(\r)$ value in that plaquette's center. 
(Notice that the possible $z(\r)$ values on a particular site can only
differ from the initial value by a multiple of 4; the values of $z(\r) \pmod 4$
define four fixed, square sublattices.\cite{zheng}). 

Now let $\{p_\alpha(t)\}$ be the probability 
of being in microstate $\alpha$ at time $t$. 
The master equation states
   \begin{equation}
         {{dp_\alpha(t)}\over{dt}} =  
          \sum _{\langle \alpha \to \beta \rangle}  
         (p_\beta-p_\alpha)
           \equiv - \sum_\beta \W_{\alpha \beta} p_\beta
   \label{master}
   \end{equation}
where $\langle \alpha \to \beta \rangle$ means summing over configurations 
$\beta$ which differ 
from $\alpha$ by one dimer flip; thus

   \begin{equation}
    \W_{\alpha\beta} = F_\alpha \delta_{\alpha \beta} - 
                          {\cal A}_{\alpha\beta}
   \label{Ldef}
   \end{equation}
where $\cal A$ is the adjacency matrix (elements unity if 
$\alpha$ and $\beta$ are related by a flip, and zero otherwise), 
and $F_\alpha\equiv \sum _\beta {\cal A}_{\alpha\beta}$ 
is the number of ``flippable'' plaquettes in
configuration $\alpha$. 
These matrices are $M_0 \times M_0$, where $M_0$ is the number of microstates.
It is well known that the matrix $\W$
has nonnegative eigenvalues 
(via the Perron-Frobenius theorem, since $\W$ is 
a stochastic matrix).
The eigenvector of zero eigenvalue is $p_\alpha= 1/M_0$, 
the weight of the (equilibrium) steady state 
(which is unique, since dimer flips connect all microstates). 
Furthermore, any time correlation function in the system can be resolved into
a sum over eigenvalues $\lambda$ of $\W$, in the form 
   \begin{equation}
       \sum _\lambda c_\lambda e^{-\lambda t}
   \label{dyncorr}
   \end{equation}
Then it is normal to define the system's equilibration time 
   \begin{equation}
     \tau(L) = {\lambda'_{min}}^{-1}, 
   \label{lambdamin2}
   \end{equation}
where $\lambda'_{min}$ is the smallest nonzero relaxation rate. 
The equilibration time $\tau(L)$ of the system
is defined as the inverse of
the second smallest eigenvalue of $\W$.

\subsection{Quantum dimer model}
\label{disc-qdim}

In the quantum dimer model, the basis states
are taken to be the dimer coverings, and the 
Hamiltonian is taken to have matrix elements
  \begin{equation}
     {\cal H_{\alpha \beta}} = 
      -  {\cal A_{\alpha\beta}} + V 
       F_\alpha \delta_{\alpha\beta}.
  \label{qdimer}
  \end{equation}
The first term describes dimer flips -- like a particle quantum
``hopping'' with amplitude $t$ on the microstate graph; the second term
is a ``potential energy'' 
which penalizes each flippable plaquette.

When $V=1$, obviously 
  \begin{equation}
    {\cal H_{\alpha \beta}}  = \W_{\alpha\beta}
  \label{H-W}
  \end{equation}
Ref.~\cite{rokhs} noted that (\ref{H-W}) implies
the ground state wavefunction
of (\ref{qdimer}) 
is a superposition of all the dimer packings with equal amplitudes. 
In other words, that (quantum) wavefunction 
is proportional to the steady-state 
probabilities of the (classical) dynamics (\ref{master}). 

Here I note that (\ref{H-W}) further implies a one-to-one correspondence of
{\it all} the eigenstates of (\ref{qdimer}) to normal modes
of the master equation.
Thus the bounds derived in this paper for the slowest relaxation
rate are equally valid for the energy gap
in the quantum model, and the approximate eigenfunctions 
and eigenvalues found in Secs.~\ref{cont}  and~\ref{rwalk}
also describe the low-energy spectrum of the quantum dimer model. 

\section{CONTINUUM THEORY OF DYNAMICS}
\label{cont}

This section develops the continuum
(``coarse-grained'') version of the dynamics.
In this form the model has an easily visualized physical meaning 
and (being linear) is solvable by standard and almost trivial techniques. 
In general, the slowest modes are associated with
relaxation of the ``order parameter'' \cite{hoh,chaik};
in the dimer covering model, the height variable plays the
role of a (hidden) order parameter, so the dynamics are phrased in
terms of it.

It should be noted that this theory is general to all
height models\footnote{
In the cases that $z(\r)$ has more than one component, 
it is necessary to let $\Gamma$ be a tensor.}
The only specific information from the dimer model is the
numerical value of the elastic constant $K$ and of the height
space lattice constant $a_h$, which  are also known for 
many other height models.

\subsection {Continuum equations and Fourier modes}
\label{cont-eqs}

First I review the coarse-grained picture 
of the height dynamics.
The static free energy functional has the form
   \begin{equation}
   F= \int_{[0,L]^2} d^2 \r {K\over 2} |\nabla h(\r)|^2
   \label{freeE}
   \end{equation}
Here $h(\r)$ represents a smoothed version of $z(\r)$ and
$K$ is
the stiffness constant controlling the fluctuations of the
 ``interface''.
(From here on I assume zero net tilt of $z(\r)$ and of $h(\r)$ so these
variables satisfy periodic boundary conditions.)
The fact that the dimer model is rough (described by (\ref{freeE})
is nontrivial: 
several other height models, defined in similar ways, 
are found to be smooth\cite{rag,kondev,burt,zeng}
in which the interface turns out to be smooth, or marginal. 
The roughness is confirmed only through the 
calculation in App.~\ref{static}.

The customary dynamics for such a field theory
(see, e.g., \cite{chaik}) is formulated as a Langevin 
equation,
   \begin{equation}
    {{dh(\r)}\over{dt}} = - \Gamma
         {{\delta F(\{ h(\cdot) \} )}\over{\delta h(\r)}}
               + \zeta (\r,t)
   \label{langevin}
   \end{equation}
Here $\Gamma$ is the kinetic (damping) constant measuring
the linear-response
to the force $\delta F/\delta h(\r)$, and $\zeta(\r,t)$ is
a random source of Gaussian noise, uncorrelated 
in space or in time.
In order that Eq.~(\ref{langevin}) have for its steady state 
$\exp(-F)$ with $F$ given by Eq.~(\ref{freeE}), the usual condition
   \begin{equation}
     \langle \zeta(\r,t) \zeta(\r',t') \rangle = 2 \Gamma
     \delta(\r-\r') \delta (t-t')
   \label{noise}
   \end{equation}
must be satisfied.

In the context of models of real (rough) interfaces of cry
stals,
Eq.~(\ref{langevin}) is known as the 
Edwards-Wilkinson process~\cite{edw}.
There is a large literature on more elaborate
equations of this form (usually with nonlinear terms)
\cite{bar}.


To make plausible the assumption of uncorrelated 
noise, one must consider
the action of the microscopic dynamics on the microscopic
heights
$z(\r)$. An elementary dimer flip changes $z(\r)$ on just
one plaquette by
$\pm \Delta z \equiv \pm 4$,
and the next dimer flip occurs at another random place.
Thus the net height is not conserved, the change is local,
  and
uncorrelated in time, which are modeled by the identical
properties of the Langevin noise in (\ref{noise}).


We can Fourier transform the above equations,
since periodic boundary conditions
maintain translational symmetry.
My short-distance cutoff prescription for this
continuum field theory shall be that the 
fields' Fourier transforms
have support only in the Brillouin zone $(-\pi,\pi)^2$;
in other words, the only allowed $\q$ values
are those which are defined for the microscopic 
lattice model (see Sec.~\ref{micro}). 
Then (\ref{langevin}) becomes
   \begin{equation}
      {{d \th_\q }\over{d t}} =
      -\Gamma K |\q|^2 \th_\q  + {\tilde \zeta_\q(t)}
   \label{qlangevin}
   \end{equation}
with  Gaussian noise
   \begin{equation}
     \langle \tilde \zeta_\q(t)^* 
             \tilde \zeta _\q'(t') \rangle =
      2 \Gamma  \delta_{\q,\q'} \delta (t-t').
   \label{qnoise}
   \end{equation}
(Note that $\delta_{\q,\q'}$ is the 
discrete $\delta$-function, appropriate
to the discrete lattice of wavevectors corresponding 
to periodic boundary
conditions.)
Thus the different Fourier components are decoupled in 
(\ref{qlangevin}).
For each of them (except for $\q=0$), (\ref{qlangevin}) is a 
one-dimensional Langevin equation with a restoring force. 

\subsection{Dynamic scaling}

The correlation function of the heights is
easy to derive from the Langevin equation: it is
   \begin{equation}
   \langle \th_\q(0) \th_{-\q}(t) \rangle =  {1\over {K|\q|^2}} e^{-\lambda_h(\q) t},
   \label{hdecay}
   \end{equation}
where the relaxation rate is 
   \begin{equation}
   \lambda_{h} (\q) = \Gamma  K |\q|^2
   \label{lambdah}
   \end{equation}

Hohenberg and Halperin argued\cite{hoh}, 
that the dynamic exponent is best 
defined by the relation between
relaxation rates and wavevectors. 
Then (\ref{lambdah}) implies $z=2$. 
As a corollary to (\ref{lambdah}),
the smallest relaxation rate of a Fourier mode
corresponds to the
smallest nonzero wavevector $\q_{min}=(2\pi/L,0)$
i.e.
   \begin{equation}
   \lambda_h (\q_{min}) = 
       (4 \pi^2 \Gamma K) L^{-2}
   \label{lambdahmin}
   \end{equation}

Random tilings, with vertices not constrained to lie on a
periodic lattice,
are studied as models of quasicrystals\cite{hen91}. These
too can be
mapped to effective interfaces $z(\r)$.
(A complication,
unimportant for the present discussion, 
is that the 
height function $z(\r)$ or $h(\r)$ has
two or more components in the quasicrystal cases.)
In one case of a quasicrystal random tiling 
(in three spatial dimensions), 
(\ref{lambdah})
was confirmed by simulation \cite{shaw}.
After $z(\r,t)$ was constructed, 
the data were numerically Fourier transformed
to give $\tz_\q(t)$ at selected (small) wavevectors $\q$;
for $\q$  small, 
that is essentially $\th_\q(t)$.
Then the time correlations 
$\langle \tz_\q(0) \tz_\q(t) \rangle$
were fitted to 
the form (\ref{hdecay})
and a plot of $\lambda_h(\q)$ versus $\q$ revealed the behavior
(\ref{lambdah}). 
A similar method\cite{zeng} was used in a study
of the Ising antiferromagnet of general spin at $T=0$ on 
the triangular lattice, which also has a height representation. 

Ref.~\onlinecite{propp}
simulated the random dimer model with 
dimer-flip dynamics; however 
rather than periodic boundary conditions, that work
used ``Aztec diamond''
boundary conditions\cite{xpropp}
such that $h(\r)$ is fixed (and spatially nonconstant)
along the edges. 
The continuum equations below would still predict 
$\tau(L) \sim L ^2$ in that geometry, as was
observed in the simulations.\cite{propp}
Note, though, that they define $\tau(L)$ without
the benefit of Fourier analysis, as the mean time it takes 
two (initially independent) replicas of the system to coincide, 
when evolved using identical random numbers sequences. 

\subsection{Fokker-Planck mode spectrum}
\label{FP}
\label{FP-fourier}

Just as the discrete stochastic dynamics implies (\ref{master}), 
the continuum stochastic dynamics
(\ref{langevin}) implies 
the familiar Fokker-Planck
equation for the evolution of the probability density
$P( \{ \th_\q(t) \} )$:\footnote{
The customary abuse of notation is committed in which $h_\q$ {\it appears}
to be manipulated as if it were a real variable; the
convention $dh^*/dh =0$ justifies the 
manipulations in (\ref{fokker}) -- (\ref{Psi}).}
  \begin{eqnarray}
      { d \over {dt}}  P (\{ \th_\q(t) \} )  &=&
      \W_h \{ P( \{ \th_\q(t) \} ) \} \nonumber \\
    \Gamma && {d\over {d\th_\q}} 
      \sum _\q \big( {d\over {d\th_{-\q}}} +K |\q|^2 \th_\q  \big) 
      P( \{ \th_\q(t) \} )
  \label{fokker}
  \end{eqnarray}


A physicist analyzing the Edwards-Wilkinson dynamics (\ref{langevin})
would usually have stopped at (\ref{hdecay}). 
That does indeed represent the slowest
Fourier mode at each wavevector, but there are many more eigenmodes of 
the general equation (\ref{fokker}). 
These modes are worth computing because (i) they 
permit computation of more general 
correlation functions than (\ref{hdecay})
(ii) the analytic form of the modes, derived below, 
might inspire improvements
on the variational eigenfunctions used in
Sec.~\ref{micro}, and 
(iii) these modes correspond to excited states in the
quantum dimer model at its critical point 
(see Subsec.~\ref{disc-qdim}, above). 

The unique zero eigenvalue of Eq.~(\ref{fokker}), corresponds of course to
to the Boltzmann distribution which is a Gaussian:
  \begin{equation}
      P_0 (\{ \th_\q(t) \} )
      \equiv \exp(- \sum _{\q\neq 0} {1\over 2} K|\q|^2 \th_\q \th_{-\q})
  \label{eigfcn}
  \end{equation}

To construct all the other eigenfunctions, it is convenient to write 
      $\Psi (\{ \th_\q(t) \} ) \equiv
      P (\{ \th_\q(t) \} )
          P_0 (\{ \th_\q(t) \} )^{-1/2} $
so that the time evolution operator of
$\Psi( \{ \th_\q(t) \} )$ is Hermitian:
  \begin{equation}
      { d \over {dt}}  \Psi (\{ \th_\q(t) \} )
      = - \HH_h \Psi( \{ \th_\q(t) \} )
  \end{equation}
where
  \begin{eqnarray}
      \HH_h &\equiv& \Gamma  \sum_{\q\neq 0}
      \big( - {d\over {d\th_{-\q}}} +\half K |\q|^2 \th_\q  \big) 
      \big( {d\over {d\th_{-\q}}} +\half K |\q|^2 \th_\q  \big) \nonumber\\
      &= &\sum _{\q\neq 0} \Gamma K |\q|^2 \CCD_\q \CC_\q
  \label{qHam}
  \end{eqnarray}
where the ``annihilation'' operator  is
  \begin{equation}
      \CC_\q \equiv (\half K|\q|^2)^{-1/2} 
                  \big( {d\over{d \th_{-\q}}} + \half K|\q|^2 \th_\q \big)
  \end{equation}
and the corresponding ``creation'' operator is
  \begin{equation}
      \CCD_\q \equiv (\half K|\q|^2)^{-1/2} 
                  \big( - {d\over{d \th_{\q}}} + \half K|\q|^2 \th_{-\q} \big)
  \end{equation}
Of course $\CCD_\q$ commutes with 
$\CC_{\q'}$ for all $\q'\neq \q$.
Obviously this is mathematically identical to the quantum Hamiltonian for
a set of harmonic oscillators with frequencies $\lambda_h(\q)$ given 
by (\ref{lambdah}), with a ``ground state wavefunction''
$\Psi_0=P_0^{1/2}$. 

Now we can write any other eigenfunction:
  \begin{equation}
   \Psi(\{ \th_\q \}; \{ n(\q) \} ) 
   = \prod _{\q\neq 0} (\CCD_\q)^{n(\q)} 
      \Psi_0( \{ \th_\q \}) 
  \label{Psi}
  \end{equation}
where $\{ n(\q) \}$ are any nonnegative integers, 
corresponding to the occupation numbers of the oscillators. 
When translated back in terms of
      $P( \{ \th_\q(t) \} )$, 
we see that each eigenstate
is a product of 
      $P_0( \{ \th_\q(t) \} )$, 
times polynomials in 
$\{ (\half K|\q|^2 )^{1/2} \th_\q \}$. 
For the ``elementary excitation'',  in which $n(\q)=1$ 
for one wavevector and zero for all the others), this
polynomial is exactly $\th_\q$; this explains why the correlation
function (\ref{hdecay}) sees only that one eigenmode.

The eigenvalue  corresponding to (\ref{Psi}) is
  \begin{equation}
      \lambda_{tot}(\{ n(\q) \}) 
       = \sum _{\q\neq 0} n(\q) \lambda_h(\q)
  \label{lambdatot}
  \end{equation}
corresponding to the total energy of the quantum oscillators. 
The net wavevector is
    \begin{equation}
         \q_{tot} = \sum _{\q\neq 0} n(\q) \q .
    \label{qtot}
    \end{equation}
Notice the many degeneracies resulting from the fact that 
$\lambda(\q) = \lambda(-\q)$: not merely the degeneracies 
due to global symmetries such as $\q \to -\q$, but less trivial
degeneracies such as the one  between the mode with $n(\q)=n(-\q)=1$ and
the one with $n(\q)=2, n(-\q)=0$. 

\section{HEIGHT-SHIFT MODE}
\label{rwalk}

Besides the translations in real space, 
a height model (as I define it) has the additional symmetry
of translations in ``height space''
(the target space of $z(\r)$ and $h(\r)$). 
Correspondingly we will find another kind of slow mode
(in a finite system), 
corresponding to a random walk of the mean height.
which I will call the ``height-shift mode''.
It will turn out to be the slowest mode. 

Consider the average height
   \begin{equation}
          \hbar (t) \equiv  N^{-1} \int_{[0,L]^2} d^2\r ~h(\r,t)
   \label{hbar}
   \end{equation}
i.e. $N^{-1/2} \th_0(t)$
(the factor $N^{1/2}$ comes from my
normalization convention for Fourier transforms.)
Of course, this is just the $\q=0$ mode that was excluded
from in the preceding section (e.g. in (\ref{Psi}) and
(\ref{lambdatot})). 
The Langevin equation
(\ref{langevin}) with $|\q|=0$
tells us  $\th_0(t)$ simply executes a Gaussian  
random walk.  
\footnote{
This behavior (and the assertions deduced from it)
are valid only while the height model is
in a ``rough'' phase. One motivation for understanding the
relaxation modes (or quantum energies)
is that they might serve as a diagnostic to distinguish 
rough and smooth phases in simulation results.}

When $h(\r)$ describes a genuine interface, 
states with different $\hbar$ are all distinct, and
the distribution of $\hbar$ simply spreads diffusively
without ever reaching a steady state. 
However, in the dimer model (and all other
``height models''\cite{blot,kond,rag,zeng,burt})
the height map is one-to-many:
here a global shift of $z(\r)\to z(\r)+4$ describes
exactly the same dimer configuration, so the image space 
of $h(\r)$ should be considered a circle of diameter 4. 
Thus the distribution function, which begins sharply peaked at
a particular value of $\hbar$, will evolve to a 
uniform distribution at some rate.

To make this mode more concrete, it may help to compare with the
behavior of a spin operator as seen in App.~\ref{spincorr}.
A system with a height distribution peaked at (say) 1 and 3 has more dimers
in one orientation than in the other. As this $n(0)=2$ mode decays, 
this polarization of orientations will decay; thus these modes
have quite real physical meanings. 


The random walk behavior (analog of (\ref{hdecay})) is
\begin{equation}
      \langle |\hbar(t)-\hbar(0)|^2 \rangle = D(N) t
   \end{equation}
with a diffusion constant
   \begin{equation}
       D(N)=  {{2 \Gamma}\over N}.
   \label{diffconst}
   \end{equation}
It is interesting that although the continuum theory cannot provide the 
numerical value
of the coefficient $\Gamma$, it does predict 
an exact ratio
(in the limit $N = L^2 \to \infty$) between  
the relaxation rate of the
modulation $h(\q_{min},t)$, and the diffusion rate of
the wandering of $\hbar(t)$.

\subsection {Fokker-Planck modes of $\hbar(t)$}

\label {FP-rwalk}


The  Fokker-Planck equation for $\hbar$ is simply the diffusion
equation; its eigenfunctions are simply plane waves as a function of
$\hbar$, 
     \begin{equation}
         \psi_Q(\hbar) = \exp(iQ \hbar)
    \label{psihbar}
    \end{equation}
and the corresponding eigenvalue is
    \begin{equation}
         \lambdahbar(Q) = {1\over 2} D(N) Q^2
    \label{lambdahbar}
    \end{equation}
with $D(N)$ given by (\ref{diffconst}).
If $\hbar$ were diffusing on a line, then 
any $Q$ would be valid in (\ref{lambdahbar})
giving a continuum of eigenvalues.

But, as noted above, a global shift $z(\r) \to z(\r)+4$
of a microstate corresponds to the same microstate.
Thus, the only modes which can correspond to 
modes in the microscopic model
are those periodic under $h(\r) \to h(\r)+4$;
i.e.
    \begin{equation}
        Q = n(0) {\pi\over 2}
       \label{Qvalues}
    \end{equation}
for any integer $n(0)$.
and the smallest such eigenvalue is 
    \begin{equation}
         \lambdahbar(\pi/2) = {{\pi^2 \Gamma}\over {4N}} 
    \label{lambdahbarmin}
    \end{equation}

Thus the {\it complete} set of eigenfunctions are
    \begin{equation}
          \Psi(\{ \th_\q \}; \{ n(\q), \q\neq 0 \} ) \psi_{n(0)\pi/2}(\hbar)
    \label{Psigrand}
    \end{equation}
a product of (\ref{Psi}) and (\ref{psihbar}), 
with any integer $n(0)$.
The corresponding eigenvalue is 
   \begin{equation}
      \lambda_{tot}(\{ n(\q) \}, \q\neq 0) 
       + n(0)^2 \lambdahbar(\pi/2)
   \label{lambdagrand}
   \end{equation}
with the first term from (\ref{lambdatot}).
The eigenvalue (\ref{lambdahbarmin}) given by $n(0)=\pm 1$ and
$n(\q)=0$ otherwise, 
is in fact the overall smallest nonzero eigenvalue.\footnote{
Of course, (\ref{lambdahbarmin}) would not be 
smallest in the case of $h(\r)$ 
fixed along the boundaries, as in 
\cite{xpropp} and \cite{luby}, since that forbids
$\hbar$-wandering modes.
Notice also that, for systems of unequal length sides, 
the ratio of (\ref{lambdahmin}) to (\ref{lambdahbarmin})
is decreased by a factor $L_{min}/L_{max}$
so that for a sufficiently elongated system
the Fourier mode is the slowest one.}
Thus (\ref{lambdahbarmin}) 
is smaller than (\ref{lambdahmin}) by a factor $(16K)^{-1}$. 
From Appendix~\ref{static}, 
the exact value is $K=\pi/16$  so this ratio is 
$1/\pi$. 

Is the height-shift mode qualitatively distinct from the capillary wave modes?
Clearly there is a close connection -- 
the uniform fluctuation of the entire sample of diameter $L$ 
is much like the fluctuation of one quadrant of a system 
of diameter $2L$ with a long wavelength Fourier mode. 
I would argue there {\it is} a distinction, in that the rate 
of the $\hbar(t)$ relaxation
depend on ``internal'' details of the height model; it cannot
be inferred given only the capillary spectrum. 

For example, I have noted the fully-frustrated Ising model 
with single-spin-flip dynamics is identical to the dimer model 
{\it except} that it has
a height space period of 8 rather than 4. Thus one must replace
$\pi/2 \to \pi/4$ in (\ref{Qvalues}) and the smallest eigenvalue
(\ref{lambdahbarmin}) is smaller by a factor  $1/4$. 
(In the fully-frustrated Ising model, a shift $z\to z+4$ reverses all the
spins; thus the mode with $Q=\pm \pi/2$ is the slowest mode that is odd
in spin.)
In the other direction, I cannot rule out
the possibility that in some height model, 
the stiffness constant $K$ and
height periodicity $a_h$ might have numerical
values such that $K a_h^2 >1$, in which case
the slowest mode would be the Fourier mode. 

The relationship of the height-shift mode to the 
capillary wave modes (both of which are a consequence of 
height-shift sysmmetry), 
is reminiscent of the relationship between 
two kinds of low energy excitation in a quantum spin system
(both consequences of spin rotational symmetry.)
The height-shift mode is the analog of the change in 
total spin number (giving energies $S(S+1)/2\chi$);
the capillary mode is the analog of a spin-wave mode.

\section{Correspondence to discrete model}

The low-lying modes of the discrete model
should be well described by the ``quantum numbers'' 
$n(\q)$ of the capillary modes and height-shift modes
of continuum model. 
\footnote{
The discrete model has additional lattice symmetries
 (rotations by $\pi/2$ and reflections), which correspond to
additional ``quantum numbers''; 
it furthermore has a ``locking'' tendency (to
favor a particular value of $h(\r)$ (modulo 1). 
Discussion of these complications will be deferred to a later
publication \cite{henqdim}.}
Now, 
      $P_0( \{ \th_\q(t) \} )$
corresponds to the eigenmode of the discrete model
which has equal weight $p_\alpha^{(0)} = 1/M_0$ for every microstate. 
Thus the prescription for constructing the approximate
eigenmode of the discrete system is
   \begin{equation}
    p_{\alpha; n(\q)} \approx 
   {{ \Psi(\{ \tz_\q \}; \{ n(\q) \} )  \psi_{n(0)\pi/2}(\zbar)}   \over
      {\Psi_0( \{ \tz_\q \})}}
   \label{discretemode}	
   \end{equation}
Here $z_\q$ is considered an implicit
function of the discrete state label $\alpha$. 
We obtained (\ref{discretemode})
simply by replacing $\th_\q \to z_\q$, and $\hbar \to \zbar$
defined by
   \begin{equation}
 \zbar \equiv  N^{-1} \sum _{\r} z(\r,t),
   \label{zbar}
   \end{equation}
which one expects to be valid for small enough 
$n(\q)$ and $\q$. 

However, (\ref{discretemode}) does not map every mode of the
continuum equation to an approximate mode of the discrete one. 
Indeed, since each $n(\q)$ is unbounded from above, 
the continuum kernel $\W_h$ has infinitely many independent eigenmodes, 
whereas the discrete kernel $\W$ has only $M_0$ eigenmodes. 
Presumably, when $n(\q)$ is so large that 
$n(\q)|\q|^2$ is of order unity for some $\q$
the anharmonic terms that I omitted in writing (\ref{freeE})
become important and mix this mode with others. 
Thus only $\q_{tot}$ is a good ``quantum number''
for labeling the higher modes. 

The low-energy eigenstates of the quantum dimer model
(\ref{qdimer})
can be predicted from the above analysis. 
The wavefunctions should be approximated by
(\ref{Psigrand})
and the energies given by (\ref{lambdagrand}). 
These predictions could be compared\cite{henqdim} with 
the results of recent exact-diagonalization studies on
$8\times 8$ lattices\cite{leung}). 

\section {MICROSCOPIC THEORY}
\label{micro}

Now we return to the exact microscopic dynamics,
as introduced in Sec. I.
The microstates are viewed as nodes of a graph 
in an abstract space, with each possible
dimer-flip (and its inverse) forming an edge of the graph:
the dynamics described by (\ref{master})
is a random walk on this graph.
As noted already,  
the graph of microstates with tilt
zero has only one connected component.
The matrix $\W$ is not only stochastic but symmetric.

\subsection{Variational bound}
\label{micro-var}

The key idea in this section is the use of a variational
guess for the eigenfunction.
This is
mathematically equivalent to the variational bound on
the eigenenergy in the quantum dimer model
(see Subsec.~\ref{disc-qdim}.)

For any vector $\{ \phi_\alpha \}$,  the smallest eigenvalue $\lambda_{min}$ satisfies
    \begin{equation}
      \lambda_{min} \leq  \lambda_\phi \equiv
      {{ (\phi^* \W \phi)} \over {(\phi^* \phi)}}
   \label{varbound}
   \end{equation}
Here $\phi^*$ means the hermitian conjugate.
From (\ref{Ldef}), the numerator is
   \begin{equation}
      (\phi^* \W \phi) 
      = \sum _{\langle \alpha\beta \rangle}
      |\phi_\alpha-\phi_\beta|^2 
   \label{phidiff}
   \end{equation}
where $\langle \alpha \beta \rangle$ means every pair of microstates,
connected by a spin flip, is counted once.

In the two cases considered in this paper (following subsections), 
$|\phi_\alpha - \phi_\beta| \equiv |\Delta \phi|$
turns out to be the same for every flip move.
Then the sum (\ref{phidiff}) reduces to 
${1\over 2} |\Delta \phi|^2 \sum _\alpha F_\alpha$
(the $1/2$ cancels the double-counting of each graph edge), and finally to
${1\over 2} fN|\Delta \phi|^2 $ 
Here $f$ is the probability that a given plaquette 
is ``flippable'', 
i.e. $fN$ is the average ``coordination number'' of the microstate graph.
The exact value \cite{fishs} is
   \begin{equation}
      f=1/8.
   \label{fvalue}
   \end{equation}

Meanwhile, the denominator in (\ref{varbound})
is just $M_0 \langle |\phi|^2 \rangle$.
Thus finally the upper bound is
   \begin{equation}
      \lambda_\phi =
        {1\over 2} f N
         {{|\Delta \phi|^2 } \over
         {\langle |\phi|^2 \rangle}}
   \label{lambdaphi}
   \end{equation}

When applied to the entire set of eigenvalues of $\W$, 
eq.~(\ref{lambdaphi}) is not very interesting, since we
already know that $\lambda_{min}=0$ (the eigenvalue of the
steady state). However, the variational argument (and every equation
in this subsection) 
is also valid when restricted to
a {\it subspace} orthogonal to that of the ground state.
Then, the bound (\ref{lambdaphi}) can be useful since the $\lambda_{min}$ of such a subspace is usually positive.\footnote
  {In subsequent subsections, labels will be attached to $\lambda_{min}$ to
  indicate which subspace they belong with, but the labels have been omitted
  in the general argument here.}
Indeed, 
the overall smallest nonzero eigenvalue -- the goal of this paper 
(see (\ref{lambdamin2})) --
is expected to be the $\lambda_{min}$ of one of these subspaces.

The above variational argument was first presented \cite{halp}
with a physical interpretation in terms of the
normalized dynamic correlation function
   \begin{equation}
       C_\phi(t) \equiv
         {{\langle \phi(0)^* \phi(t) \rangle} \over
         {\langle |\phi|^2 \rangle}}
   \label{Cphi}
   \end{equation}
Then $C_\phi(t) = 1 - {1\over 2}
{{\langle |\phi(0) - \phi(t)|^2 \rangle} /
{\langle |\phi|^2 \rangle}}$; but at short times,
$\langle |\phi(0) - \phi(t)|^2 \rangle \cong
\langle |\Delta \phi|^2 \rangle f N t$, since there
are on average $fN$ independent places
where a flip could occur. Thus the upper bound on $\lambda_{min}$ 
can be rewritten as the {\it initial} decay rate
of the correlation function,\cite{halp}
   \begin{equation}
       \lambda_\phi = - {{dC_\phi(t)}\over {dt}}\big|_{t=0
}
   \label{dCdt}
   \end{equation}
Ref. \cite{halp} (and similarly \cite{lisok}) applied (\ref{dCdt})
to bounding the dynamic critical exponent by a function of the
static exponents (for a system with nontrivial exponents). 

In some circumstances, $\lambda_\phi$ could be a good 
estimate of $\lambda_{min}$. It would be exact if
subsequent steps in $\phi$ were uncorrelated.
(We assume our variational
$\phi_\alpha$  always has nonzero projection onto
the slowest mode.)
Then $\phi(t)$ performs a random walk, and
$C_\phi(t) = \exp (-\lambda_\phi t)$, a pure
exponential decay. 
In light of (\ref{dyncorr}), in which coefficients allowed by symmetry are
expected to be generically positive, we would obtain $\lambda_{min}=\lambda_\phi$.

However, dynamics actually adopted (in eq.~(\ref{master})
is such that the steps are manifestly anticorrelated in time.
For, if a plaquette flips,
it must undergo the reverse flip the next time it flips,
unless its configuration has been shuffled in the meantime
by flips of the adjoining plaquettes.
So one would expect (but did not prove!)
that $\lambda_{min} < \lambda_\phi$  as a strict inequality. 

\subsection{Fourier modes}
\label{micro-fourier}

First we review the use of basis states with definite
$\q$ vectors. 
Consider the action of a translation of $\r$;
this induces a permutation of microstates which 
manifestly commutes with $\W$.
Thus all eigenvectors must have a definite wavevector $\q$
(or can be chosen thus);
that is, a translation by $\r$ simply induces a multiplication
by $\exp(i\q\cdot\r)$ of the eigenvector. 
Furthermore $\W$ must have zero matrix elements between 
vectors with different $\q$; thus
$\W$ becomes block-diagonal
in the new basis of states with definite $\q$.
Hence (\ref{varbound})
is valid for the smallest eigenvalue $\lambda_{min}(\q)$
in each block of definite $\q$; that is, if
$\{ \phi_\alpha \}$ has wavevector $\q$, then
   \begin{equation}
      \lambda_{min}(\q) \leq
      {{ (\phi^* \W \phi)} \over {(\phi^* \phi)}}
   \label{varboundq}
   \end{equation}

A suitable variational state-space vector
(for $\q\neq 0$) is
suggested by the ``elementary excitation''
eigenfunctions of Subsec.~\ref{FP-fourier};
they consisted of the steady-state eigenfunction $\Psi_0$
multiplied by $\th_q$. The microscopic analog of
the gaussian $\Psi_0$ is the equal-weighted eigenfunction of $\W$, 
hence we choose
   \begin{equation}
       \phi^{(\q)} _\alpha \equiv \tz_\q(\alpha)
   \label{varphi}
   \end{equation}
(Here $\tz_\q(\alpha)$ means take the configuration $z(\r)$
 corresponding
to microstate $\alpha$ and Fourier transform it for 
wavevector $\q$).

If $\alpha, \beta$ are related by one dimer flip
on the plaquette at $\r_f$,
the corresponding configurations of $z(\r)$ 
differ only by
$\pm \Delta z \equiv \pm 4$ at $\r_f$.
Consequently $|\Delta \phi |^2 = |\Delta z|^2 /N$.
On the other hand, the denominator of (\ref{varbound}) is exactly
   \begin{equation}
        \sum_\alpha |\tz_\q(\alpha)|^2 = M_0 \langle |\tz(
\q)|^2 \rangle
   \label{denombound}
   \end{equation}
(recall each microstate is weighted equally).
Thus the variational bound has the form
    \begin{equation}
         \lambda_{min}(\q) \le 8 f/ \langle |\tz(\q)|^2 \rangle
    \label{vartz}
    \end{equation}
Finally, we know $\langle |\tz(\q)|^2 \rangle \cong 1/K|
\q|^2$
at small $\q$, as derived in Appendix~\ref{static}. 
Eq.~(\ref{tzfluc}).  This gives the result
(for small $\q$)
    \begin{equation}
         \lambda_{min}(\q) \le 8 f K |\q|^2.
    \label{varfinal}
    \end{equation}
If we assume (\ref{lambdah}), then  (\ref{varfinal}) gives a bound
on the kinetic coefficient:
    \begin{equation}
         \Gamma \le {1\over 2} (\Delta z)^2 f \equiv 8f 
    \label{gammabound}
    \end{equation}

This approach -- constructing a variational vector 
from the Fourier transform of a local operator --  is equivalent to
the ``single-mode approximation'' used for quantum many-body systems 
such as superfluid helium \cite{feynm}. 
Indeed, \onlinecite{rokhs} already noted (in the context of the
quantum dimer model) that an exact upper bound on the eigenvalue
is implied. 
Their proposed variational wavefunction is
the Fourier transform of 
an dimer density operator $n_\tau(\r)$.
In fact, this is actually just $\Delta\tau z(\r)$
(difference operator in the direction $\tau=x,y$); thus their choice
differs from (\ref{varphi}) only by a derivative.
They did not identify the excitations as capillary waves, 
but derived the exponent $z=2$ using the correlations
of \onlinecite{fishs}. 
\footnote{Caution: the correlation function
exponent entering this calculation is always exactly 2, even 
when the spin operator exponents decay by exponents $\eta\neq 2$
(see Appendix~\ref{spincorr}).
That situation arises in the other height models, or in
the dimer model when confugurations are weighted unequally.}


\subsection {Height-shift mode} 
\label{micro-rwalk}

The variational trick also works for the random
walk behavior of (\ref{diffconst}).
Obviously the microscopic analog of (\ref{hbar}) is
$\zbar (t)$ defined by (\ref{zbar}). 
Then the $\q=0$ modes can be further block-diagonalized
into modes with particular wavevectors $Q$  in height space, 
as defined in Subsec.~\ref{FP-rwalk}, hence the variational bound is
valid within each such block (to be labeled ``$(0;Q)$''). 

In analogy to (\ref{psihbar}), we use the variational vector
   \begin{equation}
    \phi^{(0;Q)} _\alpha \equiv e^{iQ \tz(\alpha)}
      \label{phizbar}
   \end{equation}
In this case, $|\Delta \phi| = |\sin(Q \Delta z/2N)|$ for any state 
$\alpha$, with $\Delta z = \pm 4$ the same as before, 
and obviously $|\phi_\alpha|=1$ in any state.
So from (\ref{lambdaphi}) we obtain the bound
$\lambda_{min}(0;Q) \le  {1\over 2} f N \sin^2(4Q/N)$, 
or
   \begin{equation}
       \lambda_{min}(0;Q) \le  8 f  (\pi/2)^2 /N 
      \label{varzbar}
   \end{equation}
for large $N$. 

Just like the results (\ref{lambdah}) and (\ref{lambdahbar})
for the two respective kinds of mode in the continuum model, 
the best variational upper  bounds for slow relaxation rates
in the microscopic model correspond respectively to
$|\q|=|\q_{min}|=2\pi/L$ in (\ref{varfinal}) or to $Q=\pi/2$ in (\ref{varzbar}).
Either kind of bound implies a lower bound on the
relaxation time $\tau$ which is $O(L^2)$. 
The better bound comes from (39), and after substituting the
value of $f$ (\ref{fvalue}) gives
   \begin{equation}
       \tau  \ge   (2/\pi)^2 N .
      \label{vartau}
   \end{equation}

It is interesting to note that (\ref{varzbar}), when compared to
(\ref{diffconst}) and (\ref{lambdahbar}), gives the same bound on
$\Gamma$ as Eq.~(\ref{gammabound}) derived from Fourier modes.
This might be taken as an approximation for $\Gamma$, which 
amounts (as already noted) to  
neglecting the anticorrelation of flips. 
Such an approximation for $\Gamma$ 
was made in a random tiling quasicrystal model. (See
Sec. IV A and footnote 36 of Ref.~\onlinecite{shaw}).
This turned out\cite{shaw} to overestimate the true
$\Gamma$ (estimated from the simulation) by about 50\%.

\section {DISCUSSION}
\label{disc}

\subsection{Summary}
\label{disc-summary}

The results in Secs.~\ref{cont} and \ref{micro}
exemplify the fruitfulness of Fourier analysis
in models with translational symmetry;
indeed two types of Fourier transform were used.
For capillary wave modes (Sec.~\ref{cont}, and
subsec.~\ref{micro-fourier}), the Fourier transform
operates with respect to translations in physical space.
For height-shift modes
(Sec.~\ref{rwalk} and Subsec.~\ref{micro-rwalk}),
it operates on translations in ``height space''.
For each type of mode, two arguments have been given, one based on a
coarse-grained field theory (Secs.~\ref{cont} and~\ref{rwalk})
and the other based on exact bounds (Sec.~\ref{micro}),
which indicate the the longest 
relaxation time scales with system size as $L^2$.

The only rigorous consequence of the
present argument is a {\it lower}
bound (\ref{vartau}) on the longest relaxation time.
For computational purposes one prefers, of course,
an {\it upper} bound on the time needed 
to equilibrate the system. Towards this 
end, the present calculation merely
suggests the possible usefulness of Fourier analysis
in such a demonstration, and warns that the
slowest mode is not a Fourier modulation at all
but the ``$\zbar$ diffusion'' (see Subsec.~\ref{FP-rwalk}).

\subsection{Computational physics}
\label{disc-comput}

This study is relevant to the dynamics of other spin models. 
In particular, its results apply directly to the critical dynamics
of the fully-frustrated Ising model on the square lattice
since that model (at its critical temperature, $T_c=0$) 
maps to the dimer covering \cite{vill}.
The single-spin-flip dynamics in the spin 
model maps to the dimer-flip dynamics used here. 

The derivations could be trivially adapted to
the dimer covering of the honeycomb lattice and 
hence to the critical ($T_c=0$) dynamics 
of the triangular  Ising antiferromagnet
(equivalent to the dimer covering \cite{blot}). 
The behavior argued here can be expected in
other spin models which have height representation 
and single-spin update rules, e.g.
the $T=0$ three-state Potts antiferromagnet on the square
lattice (equivalent to the
6-vertex model \cite{liebwu} and hence the BCSOS model).
In those cases, however, no exact results or rigorous bounds for the 
static fluctuations  are available to replace Appendix~\ref{static}.

Understanding the local dynamics addressed here is also a 
prerequisite to addressing
the dynamics of height models endowed with loop-update or
cluster-update rules,
including
fully-frustrated Ising models\cite{kand}, the three-state
Potts antiferromagnet on the square lattice
\cite{wang},  the BCSOS model\cite{ev},
and coloring models\cite{kond}.\footnote 
{It turns out that all the algorithms cited have a simple
action on the height variables, thus the existence of a height
map may be a prerequisite to the possibility of accelerating 
such models.}
More speculatively, it would be interesting to check whether there
is any connection between the Swendsen-Wang \cite{swends} algorithm
and the height representation in the case of the ferromagnetic 4-state
Potts model, the partition function of which can be mapped  to that
of a height model\cite{nienh}.

Incidentally, the static correlation exponents of height models have
been obtained from Monte Carlo simulations
much more accurately via Fourier analysis of the heights, than via
direct measurement.  
It seems likely that the same is true for dynamic measurements;
this might help in evaluating the dynamic exponent for 
cluster-acceleration algorithms (see Ref.~\onlinecite{codd})

Somewhat analogous to the loop-update rules
are the ``zipper moves''\cite{hen91,ox}
of certain random tiling quasicrystal models, such as
the (two-dimensional) square-triangle random tiling.
In that case, the correlation time was measured to be
of $O(1)$ update move per site\cite{ox}; 
however each update move
involved $O(L^2)$ sites, so the scaling of the
net correlation time was $O(N^2)$ just as for the
local dimer flip dynamics treated in the present paper.

The height field is the hidden order parameter 
in the present model (or of height models in general)
and hence is the proper way to approach the
coarse-grained, long time dynamics. 
Note that experience has shown that
the {\it static} exponents
are extracted much more efficiently from
$\langle |h(\q)|^2\rangle$ 
than from spin-spin correlations 
(using the same simulations)~\cite{kond,rag,burt}.
Therefore, I suggest numerical studies of the 
dynamics ought to
extract estimates of the correlation times 
directly from the height fields rather than 
from spin-spin correlations.

\acknowledgments

I am grateful to D. Randall, 
and P.~W.~Leung for stimulating discussions, and
to C.~Zeng, J.~Kondev, and J.~Propp for helpful comments on the manuscript. 
I thank the Institute for Advanced Study for hospitality.
This work was supported by NSF grant DMR-9612304. 

\appendix

\section{Correlations of spin operators}
\label{spincorr}

This appendix shows that a standard order parameter $m(\r)$ of 
dimer coverings\cite{zheng,rokhs} 
reduces to a function of the heights, 
and correspondingly its asymptotic 
correlations are a function of those of the heights.
(This works for the standard order parameters of all
height models; 
the derivation for equal-time correlations
is discussed at greater length in 
Refs.~\onlinecite{blot,kond,rag,zeng,burt}.)

Each site $\r=(x,y)$ of the direct (not dual lattice),
is connected to one other site $\r'=(x',y')$ by a dimer., 
Define $m(\r)=(-1)^{x+x'+1}$ ($(-1)^{y+y'+1}i$) for dimers
running in the $\pm x$ ($\pm y$) direction.\footnote{ 
One could equivalently consider $m(\r)$ as living on the 
{\it occupied bonds}, since $m(\r)$ always takes the same value at 
either end of a dimer.}
Then (with the arbitrary constant offset of $z(\r)$ fixed
as in Fig.~\ref{figure})
   \begin{equation}
       m (\r,t) = e^{ i {\pi\over 2} (\hbar(\r,t) +{3/2}) }
   \label{eta}
   \end{equation}
where $\hbar(\r,t)$, defined on direct sites, is the
average of the four $z(\r,t)$ values on the surrounding dual sites. 

Then 
   \begin{equation}
       G(\r,t) \equiv \langle m (0,0) m(\r,t) \rangle 
       \approx \exp [ -\half \left({\pi\over 2}\right)^2 
                 C(\r,t)]
   \end{equation}
where 
   \begin{equation}
            C(\r,t)\equiv 
                 \langle [\hbar(\r,t) -\hbar(0,t)]^2\rangle
   \end{equation}
Using Fourier transformation with eqs.~(\ref{hdecay}) and 
(\ref{lambdah}), one obtains
   \begin{equation}
            C(\r,t)=
              \int {{d^\q}\over{2\pi}} ~
              2[1-e^{-\Gamma K |\q|^2 t} \cos (\q\cdot\r)],
   \end{equation}
a logarithmic divergence cut off at high $\q$ by the inverse
lattice constant and at small $\q$ by 
${\rm min} (r^{-1}, [\Gamma K t]^{-1/2}$. 
Thus finally the correlation function behaves as
   \begin{equation}
       G(\r,t) \approx  {1\over {r^\eta_{\pi/2}}} 
       \Phi(r/\sqrt{\Gamma Kt})
   \label{Gscaling}
   \end{equation}
where $\Phi()$ is a scaling function, such that 
$G(r,t)\sim r^{-\eta_{\pi/2}}$  or as
$G(r,t)\sim (\Gamma K t)^{-\half \eta_{\pi/2}}$ 
depending whether $r /\sqrt{\Gamma K t}$ is much less than
or much greater than unity. 
This shows that the dynamic critical exponent defined from
the standard correlation functions is indeed $z=2$. 
\footnote{Eq.~\ref{Gscaling}
was previously derived in the context of the hexatic order parameter:
see eq. (5.13) of Ref.~\onlinecite{nel}.}

\section {STATIC FLUCTUATIONS}
\label{static}

This appendix reviews what is know exactly 
about the equal-time
fluctuations of Fourier components of $z(\r)$.
This result will be an essential lemma for the proof about
$\lambda(\q)$ outlined in Sec.~\ref{micro}.

Evidently, in the continuum picture,  
the equal-time expectations
are given by
   \begin{equation}
        \langle \th_\q^* \th_{\q'} \rangle =
        \delta _{\q,\q'} { 1 \over  {K|q|^2}}
   \label{hfluc}
   \end{equation}
(Recall that $\th_\q^* \equiv \th_{-\q}$.)
This scaling is equivalent to
   \begin{equation}
        \langle |h(\r)-h(\r')|^2 \rangle =
        O(1) + {1 \over {2 \pi K}} \ln |\r-\r'|
   \end{equation}
in direct space.

In the study of ``height models''\cite{rag,kond,burt,zeng} and
random-tiling quasicrystals\cite{hen91},
it is the fundamental hypothesis that $\tz (\q)$  
has a behavior like
(\ref{hfluc}) at small $\q$.
The behavior (\ref{hfluc}) has been proven rigorously
for a very limited set of discrete models\cite{froehl}, 
and not even for the well-known BCSOS model  
of rough interfaces\cite{spenc}.
Even though the latter model is 
``exactly soluble''\cite{vanb}
via the Bethe ansatz,
such correlation functions are not known exactly
(correlation functions are notoriously difficult to
extract from the Bethe ansatz method\cite{baxt}).

However, the dimer covering model (with 2D Ising models)
belongs to the ``free-fermion'' class of exactly soluble
models \cite{fish,kast}, and using the Pfaffian approach
the dimer-dimer correlation functions can (in principle)
be written out exactly\cite{fishs}.
Later on McCoy evaluated the large-distance 
asymptotic behavior of the correlation function, for any 
orientation of the vector
between the two dimers~\cite{youngbl}.

Say that $n_x(\r)=1$ when there is a dimer connecting
$\r$ to $\r+[1,0]$
and zero otherwise;
$\delta n_x (\r) \equiv n_x(\r)-\langle n_x(\r) \rangle
\equiv n_x(\r)-\quart$,
similarly for $n_y(\r)$.

It was computed \cite{fishs} that, in the equal-weighted
ensemble of dimer coverings,
    \begin{eqnarray}
    C_{xx}(\R) \equiv 
    \langle \delta n_x(\r) \delta n_x(\r') \rangle = \nonumber \\
       -g(X,Y)^2 + g(X+1,Y)g(X-1,Y) 
   \label{Cxx}
   \end{eqnarray}
Here $\R \equiv (X,Y) \equiv \r'-\r $,
and $g(X,Y)$ is a Green's function arising from the Pfaffians \cite{fishs}.
At large $\R$, 
    \begin{equation}
         g(X,Y)  \cong
         \cases { 
             -{1\over \pi} {Y\over{R^2}},& $X$ odd, $Y$ even;\cr
             {i\over \pi} {X\over{R^2}}, & $X$ even, $Y$ odd;\cr
              0  & otherwise.\cr}
    \label{gXY}
    \end{equation}
Hence,  collecting the four cases ($X$ and $Y$ even or odd), 
    \begin{equation}
    C_{xx}(X,Y) \cong 
    {1\over {2\pi^2}} \big[
      (-1)^{X+Y} {{Y^2-X^2}\over{R^4}} +
      (-1)^X {1\over{R^2}} \big]
    \label{Cxxasymp}
    \end{equation}

Dimer correlations imply correlations in height gradients
since the definition  of $z(\r)$ 
(Sec.~\ref{intro})
is equivalent to
\begin{mathletters}
   \begin{eqnarray}
   z(\r+[\half,\half])-z(\r+[-\half, \half]) &= 
     & (-1)^{x+y} 4 \delta n_y(\r) \label{dzdx} \\
   z(\r+[\half,\half])-z(\r+[\half, -\half]) &= 
       &  (-1)^{x+y+1} 4 \delta n_x(\r)
   \label{dzdy}
   \end{eqnarray}
\end{mathletters}

Combining (\ref{Cxxasymp}) and (\ref{dzdy}), with  
   \begin{equation}
     z(\r+[\half,\half])-z(\r+[\half, -\half]) \to
     \nabla_y z(\r)
   \end{equation}
gives
\begin{mathletters}
\label{DzDz}
   \begin{equation}
       \big\langle \nabla_y z(\r) 
               \nabla_{y'} z(\r') \big\rangle
       \cong{{4^2}\over {2\pi^2}}
       \big[ {{Y^2-X^2}\over{R^4}} + (-1)^Y {1\over {R^2}}
 \big]
   \label{dzdxdzdx}
   \end{equation}
The term in (\ref{dzdxdzdx}) multiplied
by $(-1)^Y$ will be neglected since it averages to zero.
Similarly, using the 
analogous formulas for $\langle \delta n_x \delta n_y \rangle$, 
   \begin{equation}
       \big\langle \nabla_x z(\r) 
               \nabla_{y'} z(\r') \big\rangle
       \cong- {{4^2}\over {2\pi^2}}
        {{2XY}\over{R^4}}
   \label{dzdxdzdy}
   \end{equation}
and
   \begin{equation}
       \big\langle \nabla_x z(\r) 
               \nabla_{x'} z(\r') \big\rangle
       \cong-{{4^2}\over {2\pi^2}}
       \left[ {{X^2-Y^2}\over{R^4}} + (-1)^X {1\over {R^2}}
        \right]
   \label{dzdydzdy}
   \end{equation}
\end{mathletters}

The non-oscillating correlations in Eqs.~(\ref{DzDz})
have the form of dipole-dipole interactions, as if 
$\nabla z(\r)$ is a polarization.\cite{youngbl}
Integrating these equations
with respect to $\R$,
    \begin{equation}
       \langle |z(\r)-z(\r')|^2 \rangle
       \cong {\rm const} + {16\over{\pi^2}} \ln |\r-\r'|
    \end{equation}
and Fourier transforming yields the required result,
    \begin{equation}
       \langle |\tz(\q)|^2 \rangle
       \cong {16\over{\pi |\q|^2 }}.
    \label{tzfluc}
    \end{equation}
The behavior is indeed described by the continuum mode (\ref{hfluc}), 
with $K=\pi/16$.

An exact result for arrow-arrow correlations
was also obtained for a special case of the 6-vertex model
\cite{suth},
and applied to compute exactly the coefficient of the logarithmic
asymptotic behavior of $\langle |z(\r)-z(\r')|^2\rangle $
for the corresponding BCSOS model \cite{abr}.
(The BCSOS model is just the height mapping of 
the 6-vertex model).
Actually, the parameter values in that special case
make it equivalent
to a random dimer covering.
To convert the $h(\r)$ of the dimer model to
the height $h_{BCSOS}(\r)$ of a BCSOS model
(with lattice constant $\sqrt 2$),
put $h(\r)/2 = h_{BCSOS}$ on
the sublattice with even $h(\r)$ and erase $h(\r)$ on the
sublattice where it is odd.
(See also \cite{baxt2};
to  relate this to
other versions of the same mapping between
models\cite{suthpc}
reverse all the arrows pointing along vertical bonds).



\begin{figure}
\caption {(a). Random dimer packing, 
showing heights $z(\r)$ (those in
flippable plaquettes are circled). 
(b). Elementary dimer flip, showing heights.}
\label{figure}
\end{figure}

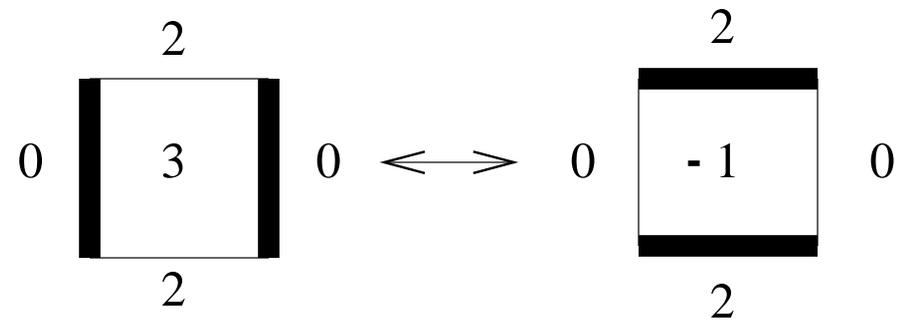
\end{document}